\documentclass[epj]{svjour}
\usepackage{graphics}
\begin{document}
\title{
The Thomas-Fermi and the Thomas-Fermi-Dirac Models in Two-Dimension-
the Effect of Strong Quantizing Magnetic Field
}
\author{Sanchari De and Somenath Chakrabarty$^\dagger$
}                     
\institute{Department of Physics, Visva-Bharati, Santiniketan, India
741 235 \\$^\dagger$E-mail:somenath.chakrabarty@visva-bharati.ac.in }
\date{Received:{\today} / Revised version: date}
\abstract{
Using Thomas-Fermi (TF) and Thomas-Fermi-Dirac (TFD) models, we
have investigated the properties of electron gas inside 
two-dimensional (2D) Wigner-Seitz (WS) cells
in presence of a strong orthogonal quantizing magnetic field.
The electron-electron Coulomb 
exchange interaction in quasi-2D case is obtained. The exact form of 
exchange term in 2D is derived making the width of the system tending 
to zero. Further, using the exchange term, the 
Thomas-Fermi-Dirac equation in 2D is established. It has been observed
that only the ionized WS cell can have finite
radius in the Thomas-Fermi model, even in presence of a strong quantizing
magnetic field. On the other hand, in the Thomas-Fermi-Dirac model a 
neutral WS cell can have finite radius.
     } 
\maketitle
\section{Introduction}
Using Fermi statistics, a formalism was developed by L.H. Thomas and E. Fermi to obtain the charge distribution and
also the distribution of electric field in the extra-nuclear space
inside heavy atoms \cite{TF1,TF2}. This formalism is known as
the Thomas-Fermi model. The electrons inside the atoms are assumed to be a degenerate Fermi
gas. In this model the electron density is found to be nonuniform
inside the atom, i.e., $n_e\equiv n_e(\vec r)$, where
$\vec r$ is the radial distance vector of the point from the nucleus
situated at the center of the atom. The nucleus at the centre has
been assumed to be a point particle. The electric
potential $\phi (\vec r)$ and the corresponding electric field vector $\vec E(\vec r)$ within the atom also vary with the radial 
vector. The electron density has been observed to be a smoothly
varying function of radial coordinate $r$, 
instead of having peaks according to shell model. 
Where the atoms are
assumed to be of spherically symmetric in shape. 
The model was successful in predicting the binding energies of the atoms \cite{LIEB}. With some suitable modification
the model has been successfully applied to molecules, solids and also to nuclei \cite{MARCH} to explain some of the
experimentally observed phenomena. The electronic shell effect was also incorporated in the model.
The model could also satisfactorily explain the thermodynamic properties of dense degenerate electron gas.
For very high density matter, the electron gas surrounding the nucleus is assumed to be enclosed in a region, called
WS cell \cite{LY}. Therefore in such situation, instead of atoms in a
solid, there are regularly spaced WS
cells, which are assumed to be charge neutral and spherically symmetric. 
There are also relativistic generalization of TF model for very high
density
electron gas and the model was found to be successful to
investigate the thermodynamic properties of such high density
degenerate electron gas \cite{REMO}.

Generalized versions of non-relativistic as well as relativistic form of Thomas-Fermi
equations in presence of strong quantizing magnetic field, when the
Landau levels for the electrons are populated have also been obtained \cite{NAG,NAG1,NAG2}.
The thermodynamic properties of matter inside the
magnetically deformed WS cells have also been investigated. In
presence of strong quantizing magnetic field, the electron pressure
becomes anisotropic inside the WS cells. As a result they will be
deformed to ellipsoidal shape from their usual spherical structure \cite{GHOSH}.

However, all these investigations are associated with three
dimensional degenerate electron gas, enclosed inside  WS
cells. There are only a few reported results on the study of two
dimensional electron gas using Thomas-Fermi model \cite{R1,R2,R3}.
Further, to the best of our knowledge, no studies have been reported
on the two dimensional Thomas-Fermi model for electron gas in
presence of strong quantizing orthogonal magnetic field, in which the Landau
levels are populated for the electrons. Again, in the three
dimensional case, incorporating  electron-electron exchange
interaction, a modified form of Thomas-Fermi equation, called
the Thomas-Fermi-Dirac equation has been developed both for
non-relativistic as well as for relativistic electron gas with or without
the presence of magnetic field \cite{LY,ST,NAG,REMO,NAG1}.
Unfortunately, no such formalism has been developed in the case of two
dimension. To the best of our knowledge, the first reported result on two
dimensional Thomas-Fermi model for degenerate electron gas in absence
of magnetic field is by
Bhaduri et. el. \cite{R1} (see also \cite{R2,R3}).

Two dimensional electron gas has a lot of important applications in modern days condensed matter physics.
The electrons in 2D are constrained to move in two dimensional sheet embedded in a three dimensional space. Such
two dimensional electron gas may be realized in many semi-conductor devices \cite{R4}. There are also possibility of having
two dimensional electron gas on the surface of materials, e.g., liquid He \cite{R5}. In such system electrons are free 
to move
on the surface of liquid He but rigidly attached with the He atoms. There are also a kind of solid insulators, e.g.,
topological insulators \cite{R6}, the surface of which supports conducting states of free electrons.

During the present days the most interesting two dimensional system in condensed matter physics is graphene. 
It is  an almost ideal two dimensional material developed in the laboratory using graphite \cite{R7,R8,R9}. It has
been observed that graphene can support 2D electron gas. It has
become a topic of current interest due to a large number of application of graphene. There are also a lot of
academic interest on the theoretical investigation of graphene. In
particular, application of quantum electrodynamics in graphene  and study
of the physics of mass-less electrons or chiral electrons using two component Dirac equation. 

In this article  we have investigated the properties of electron gas enclosed in 
two-dimensional WS cells, which are embedded in a three-dimensional
space. The WS cells are assumed to
be on $x-y$ plane and the strong magnetic field is along $z$-axis. The electrons are constrained to move on $x-y$ plane.
The presence of strong magnetic field along $z$-direction makes the electron energy eigen value discrete. The
motion of the electrons on $x-y$ plane are in quantized form. This is the well known Landau quantization in 2D. The
electron energy therefore does not depend on $p_z$, the component of
momentum along $z$-direction. However, the momentum
component on $x-y$ plane changes in a discrete manner.

We have organized the various sections of this article in the following manner:
In the next section we have developed the basic formalism for two dimensional Thomas-Fermi model for degenerate
electron gas in presence of strong quantizing orthogonal
magnetic field. In section 3 we have studied the thermodynamic properties of 2D degenerate electron gas inside a WS
cell in presence of a strong quantizing magnetic field. In section 4 we have considered the electron-electron exchange 
interaction in absence of magnetic field and incorporate
this result in Thomas-Fermi condition to obtain Thomas-Fermi-Dirac equation satisfied by degenerate electron gas in
2D. In section 5 we have shown explicitly that the same technique can not be followed to obtained
Thomas-Fermi-Dirac equation for degenerate electron gas in presence of strong quatizing magnetic field. However, for
the conventional three dimensional case, one can obtain exchange energy for electrons \cite{NAG1}. 
Finally we present the conclusion of this work.
\section{Basic Formalism}
We assume that the constant external magnetic field $\vec B$ is along z-direction. In presence of this strong
external quantizing magnetic
field, the Landau levels of the electrons are populated on $x-y$
plane. Since the electrons are constrained to move on $x-y$ plane,
the momentum component along $z$-direction, $p_z=0$. The quantized form of electron energy eigen value is then given by
$\varepsilon_{n}=(n+\frac{1}{2})\hbar\omega$, where
$\omega=eB/mc$, the cyclotron frequency and $ n=0,1,2,......n_{max}$, the Landau quantum numbers, $n_{max}$ is the
upper limit of Landau quantum number which is finite if the temperature of the system is zero or less than the 
corresponding Fermi
temperature, whereas, $n_{max}=\infty$ if the temperature of the 
electron gas is greater than the Fermi temperature. 
Therefore $n_{max}$ is finite
for strongly degenerate electron gas, whereas in the non-degenerate scenario, $n_{max}$ is infinitely large. In
the strongly degenerate condition the number of electrons per unit surface area is given by 
\begin{equation}
n_e=\frac{eB}{2\pi^{2}}\left(n_{max}+1\right )
\end{equation} 
Assuming $\varepsilon_{F}=(n_{max}+\frac{1}{2})\hbar\omega $ as the Fermi
energy of the system, we have in natural units $(\hbar=c=1)$, 
\[
n_{max}=\frac {\varepsilon_{F}}{\omega}-\frac {1}{2}
\]
Then it is a little algebra to check after putting the value of
cyclotron frequency that the maximum value of
Landau quantum number decreases with the strength of 
magnetic field. It becomes zero for $B/B_c=2\varepsilon_F/m$, where
$B_c\approx 4.43\times 10^{13}$G. In the present model
calculation, there is an upper limit for the strength of magnetic,
beyond  which the Landau quantum number becomes unphysical
(negative in nature). 
Then 
\begin{equation}
n_e=\frac{eB}{\pi^{2}}\left(\frac{\varepsilon_F}{\omega}+\frac{1}{2}\right )
\end{equation}

Now for such degenerate
electron gas confined inside a 2D WS cell, at the centre of which
there is a nucleus of charge $+Ze$,
the total negative charge carried by the electrons inside the
cell is $-eN$. Therefore the WS cells are charge neutral.
Since both protons inside the nucleus and electrons within 
the WS cell carry
electric charges, the Poisson's equation satisfied by the electrostatic potential $\phi$, produced jointly by the
electrons and protons is given by 
\begin{equation}
\nabla^2\phi=2\pi en_e-2\pi Ze\delta(r)
\end{equation}
Here the nucleus is assumed to be a point object.
The Thomas-Fermi condition in this
particular situation is given by 
\begin{equation}
\mu=\varepsilon_F=\left(n_{max}(r)+\frac{1}{2}\right )\omega-e\phi(r)={\rm{constant}}
\end{equation} 
which gives 
\begin{equation}
n_{max}(r)=\frac{\mu+e\phi(r)}{\omega}-\frac{1}{2}
\end{equation}
In the present situation for
$n_{\rm{max}}=0$, we have after substituting for $\omega$, 
$eB=2m(\mu +e\phi(r))$, gives the upper limit of the external
magnetic field, beyond which the Landau quantum number becomes
negative, which is unphysical. In terms of the critical values of
the magnetic field $B_c$ for electrons at which the Landau levels
for the electrons are populated, we have 
\[
\beta=\frac{B}{B_c}=\frac{2}{m}(\mu+e\phi(r))
\]
where $eB_c=m^2 \approx 4.4\times 10^{13}R$Gauss. Therefore the
upper limit of external magnetic field depends on the radial
distance $r$ from the centre of the Wigner-Seitz cell through the
electrostatic potential $\phi(r)$.
The number of electrons per unit surface area is then given by
\begin{equation}
n_e(r)=\frac{eB}{2\pi^2}\left(n_{max}(r)+1\right
)=\frac{eB}{2\pi^2}\left(\frac{\mu+e\phi(r)}{\omega}+\frac{1}{2}\right )
\end{equation}
Since $\phi$ is a function of radial coordinate $r$, both $n_e$ and $n_{max}$ should also depend on the radial coordinate. Here
for the sake of simplicity we assume circular symmetry for the WS cells. The above mentioned variables therefore do not depend on the angular
coordinate $\theta$. Then from the above equation we have in the extra-nuclear space of WS cell
\begin{equation}
\nabla_r^2\phi=\frac{2\pi^2\omega}{e^2B}\nabla_r^2n_e
\end{equation} 
and from the Poisson's equation (eqn.(3)) we have for $r>0$
\begin{equation}
\frac{2\pi^2\omega}{e^2B}\nabla_r^2n_e=2\pi n_ee
\end{equation}
This is a second order differential equation for the electron density $n_e$.
On substituting the value of cyclotron frequency, eqn.(8) can be re-expressed as  
\begin{equation}
\nabla_r^2n_e=\frac{e^2}{\pi}mn_e(r)
\end{equation} 
Hence for the 2D case, with circular symmetry, the Poisson's equation in radial coordinate $r$ is given by
\begin{equation}
\frac{d^2n_e}{dr^2}+\frac{1}{r}\frac{dn_e}{dr}=\frac{e^2m}{\pi}n_e
\end{equation}
Defining the scaled radius parameter $x$, given by
\begin{equation}
r=\left (\frac{\pi}{e^2m}\right )^{1/2}x=bx,
\end{equation}
we have
\begin{equation}
\frac{d^2n_e}{dx^2}+\frac{1}{x}\frac{dn_e}{dx}-n_e=0 
\end{equation}
Which is the Thomas-Fermi equation in 2D in presence of a strong orthogonal quantizing magnetic field. The
form of this equation is exactly identical with that of field free case \cite{R1}, except the scaling parameter
$b=(\pi/e^2m)^{1/2}$. Surprisingly the radius of the circular type WS cell does not change
with the strength of magnetic field. However, in the usual three dimensional case, the radius of the spherical type WS cell
decreases with the strength of magnetic field, e.g., in
\cite{NAG1,NAG2}, the scaling parameter $\mu=(\pi/2e^3B)^{1/2}$.
Hence the actual radius $r=\mu x$ varies with magnetic field in the
functional form $1/B^{1/2}$. Further in the case of magnetically
deformed cylindrical type WS cell, both the radial as well as the axial parameters decrease with the strength of
magnetic field \cite{GHOSH}. Therefore in the present 2D scenario 
we may conclude that there will be no magnetic contraction of
WS cells and even in
presence of ultra strong quantizing magnetic field, there will
be no magnetic distortion to elliptical shape. Now it is trivial to show that
the general solution of eqn.(12) is given by \cite{R10}
\begin{equation}
n_e(x)=A_1 I_0(x)+B_1 K_0(x)
\end{equation}
where $A_1$ and $B_1$ are two unknown constants to be obtained from the initial and the boundary conditions and can be expressed in
terms of $x_s$, the surface value of $x$, $N$, the total number of electrons in the system and $Z$, the total positive
charge within the nucleus situated at the centre of the WS cell. From the above solution it is quite obvious
that $\phi(x)\longrightarrow \infty$ as $x\longrightarrow 0$ from the
diverging nature of the modified Bessel function $K_0(x)$. (The
modified Bessel function of second kind of order zero, $K_0(x)\sim
-\ln x~~{\rm{as}}~~ x\rightarrow 0$). This is
also true in the case of three dimensional Thomas-Fermi equation, having a singular nature at the origin. The special
technique developed by Feynman, Metropolis and Teller \cite{FMT} is used to solve the Thomas-Fermi differential
equation in the three dimensional case. 

Since the TF differential equation is exactly identical with eqn.(10)
of \cite{R1} then instead of repeating the calculations of
\cite{R1}, we shall use eqn.(17) to eqn.(20) from this article
with our changed notations and are given by 
\begin{equation}
x_s n^\prime(x_s)=\frac{(N-Z)}{2\pi b^2}
\end{equation}
The right hand side of this equation vanishes for $N=Z$, i.e., for a charge neutral two dimensional WS cell.
\begin{equation}
A_1=-B_1\frac{K_0(x_s)}{I_0(x_s)}
\end{equation}
where
\begin{equation}
B_1=\frac{Z}{2\pi b^2}
\end{equation}
It should be noted that although the physical meaning of $b$ is same
but its expression is quite differnt from \cite{R1}.
Finally
\begin{equation}
I_0(x_s)=\frac{Z}{Z-N}
\end{equation}
It is quite obvious from eqn.(17) that for $N=Z$, the radius of the WS cell is infinitely large. It is also to be noted further 
that for the finite value of the radius parameter $x$, $N$ should be less than $Z$, i.e., the cell must be in ionized state 
with net positive charge. In fig.(1) we have plotted the variation 
of $x_s$, the scaled radius
parameter with $N/Z$. From the figure it is quite obvious that $x_s$ is finite for
$N/Z<1$. Whereas for the ratio tending to unity, the scaled radial
parameter increases sharply finally it diverges at $N/Z=1$. Therefore
in 2D, in the TF case, the radius of a positively charged WS cell is finite,
whereas for the charge neutral case the radius of the WS cell
diverges. This is true even if there is stong quantizing orthogonal 
magnetic field in 2D.
\section{Thermodynamics of 2D Electron Gas}
The expressions for internal energy density and the corresponding kinetic pressure can be obtained from the first
principle following the standard books on statistical mechanics. We define the $q$-potential of Kramers in the form \cite{LSTAT}
\begin{equation}
q=\frac{PS}{kT}= \sum_i\ln[1+\exp(-\alpha^\prime-\beta\varepsilon_i)]
\end{equation}
where $P$ is the kinetic pressure, $S$ is the surface area, $T$ is the temperature of the system, $k$ is the Boltzmann
constant, $\alpha^\prime=\alpha-\beta e\phi$, $\alpha=-\mu/kT$, with $\mu$, the chemical potential of the electrons and $\beta=1/kT$. In presence of a strong
orthogonal quantizing magnetic field, which populates the electron Landau levels, the energy eigen value
corresponding to the $n$th. Landau level is given by 
$\varepsilon_n=(n+\frac{1}{2})\hbar\omega=(n+\frac{1}{2})\omega$ for $\hbar=1$.
The total kinetic energy of the electron gas is then given by 
\begin{equation}
E=\left (\frac{\partial q}{\partial\beta}\right
)_{\alpha,S}=\frac{eBS}{2\pi^2}\sum_{n=0}^\infty\frac{\varepsilon_n}{\exp[\beta(\varepsilon_n-\mu^\prime)]+1}
\end{equation} 
where $\mu^\prime=\mu +e\phi$. For degenerate case, since the Fermi distribution function reduces to unity, we have
\begin{equation}
E=\frac{eBS}{2\pi^2}\sum_{n=0}^{n_{max}}\left (n+\frac{1}{2}\right )\omega
\end{equation}
On summing over $n$ we can express the surface density of electron kinetic energy in the form
\begin{equation}
\epsilon(x)=\frac{e^2B^2}{4\pi^2m}\left (n_{max}(x)+1\right )^2
\end{equation}
The kinetic pressure can also be obtained from $q$-potential. Using the Euler summation formula (discussed in the
Appendix) the degeneracy pressure for electron gas in 2D is given by
\begin{equation}
P(x) =\frac{m}{4\pi^2}\left (\mu+e\phi(x)\right )^2 =
\frac{e^2B^2}{4\pi^2m}\left (n_{max}(x)+\frac{1}{2}\right )^2
\end{equation}

\section{Electron Exchange Energy for $B=0$ in 2D}
We shall now obtain the electron exchange energy (see \cite{LY} for
three-dimensional case). For the sake of simplicity we start with an ideal two-dimensional
system of degenerate electron gas.
In the case of a purely two-dimensional electron gas, the electron-electron two body potential is logarithmic in
nature and may be expressed in the form
\begin{equation}
V_{ee}=e^2\ln\left (\frac{\mid\vec{r}-\vec{r^\prime}\mid}{a}\right )
\end{equation}
where as stated before, $e^2$ has the dimension of energy.
In 2D, the normalized form of free electron wave functions are given by
\begin{equation}
\psi (\vec r)=\frac{1}{S^{\frac{1}{2}}}\exp(i\vec k.\vec r)
\end{equation}
Then the exchange part of interaction may be written as
\begin{eqnarray}
U_{ex}(\vec r)\psi_i(\vec r)&=&\sum_{j=1}^N\psi_j(\vec r)\int\psi_j^*(\vec r^\prime)V(\mid\vec r-\vec
r^\prime\mid)\nonumber \\ && \psi_i(\vec r^\prime) d^2r^\prime
\end{eqnarray}
which after some straight forward algebra may be expressed in the form
\begin{eqnarray}
U_{ex}(\vec r)\psi_i(\vec r)&=&
\frac{e^2}{S^{\frac{3}{2}}}\sum_j\exp(i\vec k.\vec r)\nonumber \\ 
&&\int\exp(i(\vec k-\vec
k^\prime).(\vec r^\prime-\vec r)) \nonumber \\ &&V(\mid\vec r-\vec r^\prime\mid).d^2r^\prime
\end{eqnarray}
Hence the expectation value for the electron exchange energy is given by
\begin{eqnarray}
U_{ex}&=\frac{e^2}{S} \sum_{j} \int\exp(i\vec K.\vec s) V(s)d^2s
\end{eqnarray}
where $\vec K=\vec k-\vec k^\prime$, $\vec s=\vec r-\vec r^\prime$ and $d^2s=sdsd\theta$.
Then the angular part of the integral is given by
\begin{equation}
\int_0^{2\pi}\exp(iKs\cos\theta)d\theta
\end{equation}
Now to evaluate this angular integral we use the following trivial algebraic relation 
\begin{equation}
\exp(iks\cos\theta)=\cos(ks\cos\theta)+i\sin(ks\cos\theta)
\end{equation}
and use the standard relations \cite{R10}
\begin{equation}
\cos(z\cos\theta)=J_0(z)+2\sum_{k=1}^\infty(-1)^k J_{2k}(z)\cos(2k\theta)
\end{equation}
and
\begin{equation}
\sin(z\cos)=2\sum_{k=0}^\infty(-1)^k J_{2k+1}(z)\cos(2k\theta)
\end{equation}
where $J_n(z)$ is the ordinary Bessel function of order $n$.
Then after some straight forward algebra we have the exchange interaction part
\begin{eqnarray}
U_{ex} &=& \frac{2\pi e^2}{S}\sum_j\int sdsJ_0(ks)V(s) \nonumber \\ 
&=& \frac{2\pi e^2}{(2\pi)^2}\int d^2k^\prime\int sdsJ_0(ks)V(s)
\end{eqnarray}
where we have used the relation
$\frac{1}{S}\sum_j\longrightarrow\frac{1}{(2\pi)^2}\int d^2k^\prime$.
The integral  given by eqn.(32) is obviously a diverging one for logarithmic form of $e-e$ coulomb potential as
given in eqn.(23). The
coulomb exchange energy therefore has infinite contribution
in the case of ideal 2D electron gas, which is totally unphysical in nature. Hence, 
to obtain a finite contribution for the exchange part of $e-e$ interaction energy, we therefore follow an alternative approach. Instead of an
ideal two-dimensional system, we consider a quasi two dimensional degenerate electron gas with an width $\Delta z$ 
along the third dimension and we shall make it tending to zero at the end \cite{QDOT}. The wave functions in
the $x-y$ plane is as usual are given by eqn.(24), with $\vec r$ replaced by $x$ and $y$. In the $z$-direction the wave
function is assumed to be given by 
\begin{equation}
\phi_0(z)=\frac{1}{(\Delta z)^{\frac{1}{2}}}\cos\left (\frac{\pi
z}{\Delta z}\right)
\end{equation}
Then following \cite{QDOT} (see also \cite{AKR}), we have for the limiting case $\Delta z\longrightarrow 0$,
\begin{equation}
U_{ex}=-\frac{4}{3\pi}(2\pi n_e)^{\frac{1}{2}}
\end{equation}
where $n_e=k_F^2/2\pi$.
The modified form of Thomas-Fermi condition is then given by
\begin{equation}
\mu=\frac{k_f^2}{2m}-e\phi-\frac{4}{3}\left (\frac{2}{\pi}\right )^{\frac{1}{2}}n_e^{\frac{1}{2}}
\end{equation}
Which is the Thomas-Fermi-Dirac condition in 2D scenario. Hence
\begin{equation}
k_F=\frac{4m}{3\pi}+m\left
(\frac{16}{9\pi^2}+\frac{2}{m}(\mu+e\phi)\right )^{1/2}
\end{equation} 
The other solution for Fermi momentum has been discarded for obvious reason.
Let us now substitute
\begin{equation}
\frac{8m}{9\pi^2}+(\mu+e\phi)=\frac{Ze^2}{r}\omega(r),
\end{equation}
Then we have from eqn.(36)
\begin{equation}
k_F=\frac{4\pi}{3m}+(2m)^{\frac{1}{2}}\frac{Z^{\frac{1}{2}}e}{r^{\frac{1}{2}}}\omega^{\frac{1}{2}}
\end{equation}
Hence from the Poisson's equation we can write down the Thomas-Fermi-Dirac equation in the following form
\begin{equation}
r^2\frac{d^2\omega}{dr^2}-r\frac{d\omega}{dr}+\omega=r^32me^2\left [\frac{\omega^{1/2}}{r^{1/2}}+\frac{4\pi}{3m}
\frac{1}{(2mZe^2)^\frac{1}{2}}\right ]^2
\end{equation}
Writing $r=ax$, with $x$, the dimensionless radius parameter and $a$ is an unknown constant, the above equation can
be expressed in the following form:
\begin{equation}
x^2\frac{d^2\omega}{dx}-x\frac{d\omega}{dx}+\omega=x^3\left
[\frac{\omega^{\frac{1}{2}}}{x^{\frac{1}{2}}}+\alpha\right ]^2
\end{equation}
with 
\begin{equation}
a=(2me^2)^{-\frac{1}{2}} ~~{\rm{and}}~~
\alpha=\frac{4\pi}{3m}\left [\frac{1}{Z(2me^2)^{\frac{3}{2}}}\right ]^{\frac{1}{2}}
\end{equation}
The complementary function can be obtained from the solution of the homogeneous equation
\begin{equation}
x^2\frac{d^2\omega}{dx^2}-x\frac{d\omega}{dx}+\omega=0
\end{equation}
To get an analytical solution, we put $x=\exp(z)$, with $-\infty \leq z \leq z_s$, where $z_s$ corresponds to
surface value. Hence we have
\begin{equation}
(D^2-2D+1)\omega(z)=0
\end{equation}
where $D=d/dz$. The solution of this equation, which is the complementary function is then given by
\begin{equation}
\omega=(A_0+B_0 z)\exp(z)+1
\end{equation}
where $A_0$ and $B_0$ are two unknown constants. The factor $1$ is put by hand to get the initial condition
$\omega\longrightarrow 1$ as $z \longrightarrow -\infty$ or equivalently  $x \longrightarrow 0$.
In terms of this new variable z, the full form of Thomas-Fermi-Dirac equation is then given by
\begin{equation}
(D^2-2D+1)\omega(z)=\exp(3z)\left (\omega^{\frac{1}{2}}\exp\left (-\frac{z}{2}\right )+\alpha\right )^2=F(z)
\end{equation}
To obtain the particular integral, we follow the iterative technique. On the right hand side of the above equation
we put the value of complementary function for $\omega$ which we call as the zeroth iteration term  and obtain the equation
\begin{equation}
(D^2-2D+1)\omega(z)=F_0(z)
\end{equation}
Defining $(D-1)\omega=u$, then
\begin{equation}
\frac{du}{dz}-u=F_0(z)
\end{equation}
where 
\[
F_0(z)=\exp(3z)[\{(A_0+B_0z)\exp(z)+1\}^{1/2} \exp(-z/2)+\alpha]^2
\]
Hence using the standard technique of integrating factor, which in this case is $\exp(-z)$, we have
\begin{equation}
u(z)=\exp(z)(F_1(z))
\end{equation}
where 
\begin{eqnarray}
F_1(z)&=&\int_{-\infty}^z\exp(2z^\prime)[\{(A_0+B_0z^\prime)\exp(z^\prime)\}^
{\frac{1}{2}}\nonumber \\ &&
 \exp(-z^\prime/2)+\alpha]^2 dz^\prime
\end{eqnarray}
Here to obtain the complete solution, we demand that $\exp(-z)u \longrightarrow 0$ as $z \longrightarrow -\infty$.
Using the same integrating factor technique, one can solve the equation
\begin{equation}
\frac{d\omega}{dz}-\omega=\exp(z) F_1(z)
\end{equation}
Hence the complete solution is
\begin{equation}
\omega(z)=\exp(z)(A_0+B_0z)+1+\exp(z)\int_{-\infty}^{z} F_1(z^{\prime}) dz^{\prime}
\end{equation}
Obviously $\omega(z) \longrightarrow 1$ as $z \longrightarrow -\infty$.
It can further be shown that the boundary condition in $z$-coordinate is given by 
\begin{equation}
\frac{d\omega}{dz}\mid_{z=z_s} = \omega(z)\mid_{z=z_s}
\end{equation} 
Hence we get $B_0=1/x_s$. This particular value of $B_0$ has been used 
in the complete solution as
given by eqn.(51). The constant
$A_0$ can be
obtained numerically, provided the surface value $x_s$ is known. Further, the surface value of scaled radius
parameter and the constant $A_0$ can be solved numerically
using the expression for total number of electrons, given by
\begin{equation}
N=Z=\int d^2r n_e(r)
\end{equation}
and eqn.(52). In $x$-coordinate eqn.(53) can be expressed as 
\begin{equation}
N=Z=b\int_0^{x_s}dx\left \{\frac{4\pi}{3m}b^{\frac{1}{2}}x^{\frac{1}{2}}+(2m)^{\frac{1}{2}}
(Ze^2)^{\frac{1}{2}}\omega^{\frac{1}{2}}\right \}^2
\end{equation}
In this case we have assumed that the number of electrons within the WS cell is equal to the number of protons
within the nucleus situated at the centre of the cell. The WS cells
are therefore charge neutral. As a consequence, the 
electric field at the surface vanishes exactly.
We continue the  iterative 
calculation for the particular integral, until the result converges.  
In fig.(2) we have plotted the variation of scaled surface radius
parameter with the atomic number $Z$, which is
also the number of electrons within the two-dimensional WS cell. It
is obvious that initially for the low values of $Z$, the scaled parameter
increases monotonically with $Z$ and finally for large $Z$ it saturates to
some constant value. This type of variation has not been
observed in the case of Thomas-Fermi model in 2D. Hence we may
conclude that the radius of he charge neutral WS cell is finite in
TFD model.
\section{Electron Exchange Energy for $B\ne 0$ in 2D}
In presence of a strong quantizing magnetic field the semi-analytical expression for exchange energy can be 
obtained for electrons  assuming that all of them occupy only their zeroth
Landau level. Here again we consider a quasi-2D electron gas. 
In this case the numerically fitted form of exchange
energy is given by \cite{NAG1,NAG2}
\begin{equation}
U_{ex}=\alpha[1-\exp(-\beta p_F)]
\end{equation}
with the average values for $\alpha \approx 5 MeV$ and $\beta \approx 1.5 MeV^{-1}$  
for the strength of magnetic field ranging from $10^{13}$G to $10^{15}$G. This range has special importance in
the physics of strongly magnetized stellar electron gas. In the present situation, considering the exchange part
inside a strongly magnetized neutron stars or magnetars,
the Thomas-Fermi-Dirac condition is then given by
\begin{equation}
\mu=\frac{p_F^2}{2m}-e\phi-\alpha[1-\exp(-\beta p_F)]
\end{equation}
Hence one can obtain the numerically fitted form of Fermi momentum, given by
\begin{equation}
p_F=c(\mu^*+e\phi)^{\nu}
\end{equation}
where the average values for $c \approx 0.7$ and  $\nu \approx 0.5$ are for the same range of magnetic field. 
Here $\mu^*=\mu+\alpha$.
Unfortunately, the Fermi momentum $p_F$ here is along the third dimension, i.e, along $z$ axis, which does not exist
in the present scenario. Hence we can conclude
that exchange interaction term does not contribute when a 2D electron gas is placed in a strong magnetic
field which populates only the zeroth Landau level for the electrons.
\section{Appendix}
To evaluate the kinetic pressure for degenerate electron gas in 2D in presence of strong quantizing
magnetic field, we use the Euler summation formula as given bellow
\cite{R10}
\begin{equation}
\sum_{j=0}^{\infty}f\left (j+\frac{1}{2}\right ) \approx \int_0^{\infty} f(x)dx +\frac{1}{24}f^{\prime}(0)
\end{equation}
Then from eqn.(18) the q-potential can be re-written in presence of an electrostatic potential $\phi$ in the
following form
\begin{eqnarray}
q &=& \sum_i \ln[1+\exp(-\alpha^{\prime}-\beta\varepsilon_i)]
\end{eqnarray}
where $\alpha^{\prime}= \alpha+\beta e\phi = \beta(-\mu-e\phi)$.
Hence
\begin{eqnarray}
q &=& \frac{SeB}{2\pi^2}\sum_{n=0}^{\infty}\ln \left [1+\exp\left (-\alpha^{\prime}-\beta\left (n+\frac{1}{2}\right
)\omega\right )\right ] \nonumber \\
&=& \frac{SeB}{2\pi^2}\left
[\int_0^{\infty}\ln[1+\exp(-\alpha^{\prime}-\beta x\omega)] dx\right
]\nonumber \\
&-& \frac{SeB}{2\pi^2}\left [ \beta\hbar\omega\frac{1}{\exp(\alpha^{\prime})+1}\right ]  
\end{eqnarray}
After evaluating the integral by parts, we have
\begin{equation}
q=\frac{SeB}{2\pi}\beta\omega\left [\int_0^{\infty} \frac{x dx}{\exp(\alpha^{\prime}+\beta
x\omega)+1}-\frac{1}{\exp(\alpha^{\prime})+1}\right ]
\end{equation}
For $T=0$, or $T\ll T_F$, where $T_F$ is the Fermi temperature, given by $\mu=T_F$, for $k_B=1$, the second term  
exactly vanishes, whereas in the first part, putting the  Fermi distribution $=1$, it gives
${\mu^{\prime}}^2/2\omega$, with $\mu^\prime=\mu+e\phi$.
Then substituting the value of the integral in eqn.(61), we have
\begin{equation}
q =\beta PS 
= \frac{SeB}{4\pi^2}\beta\frac{1}{\omega}\mu{^{\prime}}^2
\end{equation}
Hence 
\begin{equation}
P(r)=\frac{m}{4\pi^2}(\mu+e\phi(r))^2
\end{equation}
\section{Conclusion}
 It is quite surprising that the form of Thomas-Fermi differential
 equation in 2D in presence of strong quantizing magnetic field is
exactly identical with that of zero field case. The exchange part of electron energy does not exist in 2D
in presence of strong quantizing magnetic field. The Fermi momentum obtained in this case is along $z$-direction,
which is suppressed in ideal 2D case. 
The Thomas-Fermi-Dirac equation therefore can not be obtained for a 2D electron gas in presence of a
strong quantizing magnetic field. However, in absence of magnetic field, exchange energy can be obtained assuming a
quasi 2D structure and finally making the width along the third
dimension tending to zero. Hence one can formulate Thomas-Fermi-Dirac model in 2D scenario for an electron
gas. We have noticed that the radius of a two dimensional charge neutral WS cell is infinitely large. On the other
hand it is finite if $N/Z <1$, i.e., the cell is ionized and carrying
some effective positive charge. In usual three dimensional case also
the radius of an atom can not be finite in Thomas-Fermi model. However, in Thomas-Fermi-Dirac model the radius of a
two dimensional charge neutral WS cell is found to be finite. We have also noticed that the size of circular shape
WS cell does not depend on the strength of magnetic field, even if it is of astrophysical order. Because of two dimensional
structure, there is no distortion of circular type WS cells to elliptical form. On the other hand, in the usual
three dimensional case, because of pressure anisotropy within the WS cells, a distortion to ellipsoidal shape may
occur in presence of strong quantizing magnetic field. In the usual three dimension, the anisotropy increases with 
the increase of the strength of magnetic field. Then in the extreme case, it can be shown that a WS cell acquires a needle
like shape with its length along the direction of magnetic field.

\begin{figure*}
\resizebox{0.75\textwidth}{!}{%
  \includegraphics{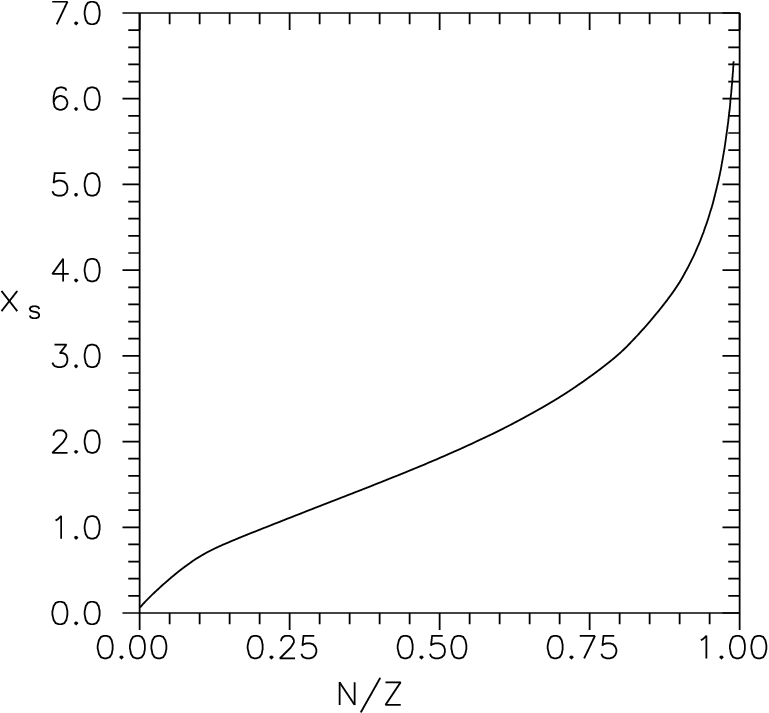}}
\caption{
Variation of $x_s$, the scaled surface value of radius parameter with the 
ratio $N/Z$
}
\label{fig:1} 
\end{figure*}
\begin{figure*}
\resizebox{0.75\textwidth}{!}{%
  \includegraphics{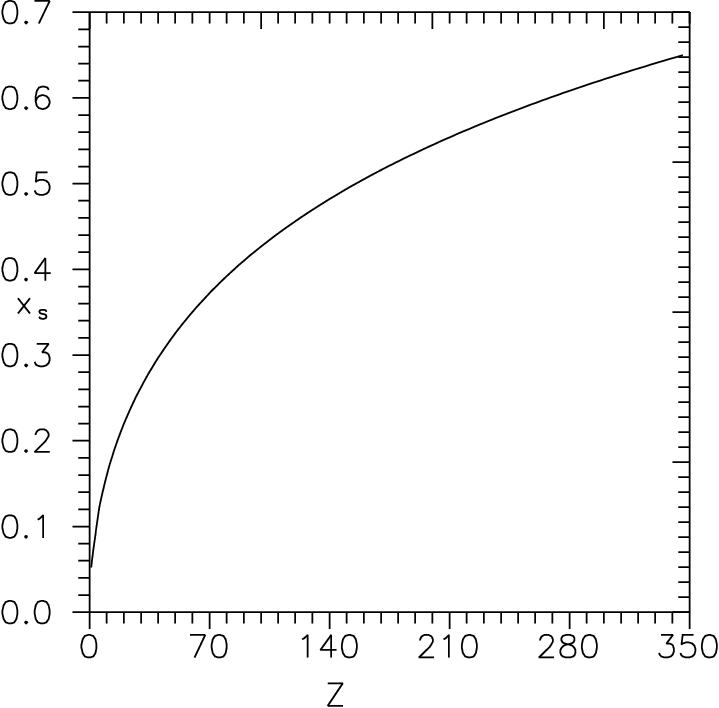}}
\caption{
Variation of $x_s$, the scaled surface value of radius parameter with $Z$
}
\label{fig:2} 
\end{figure*}
\end{document}